\documentclass[preprint2]{aastex631}

\usepackage[utf8]{inputenc}
\usepackage{gensymb}

\usepackage{graphicx}
\usepackage{natbib}
\usepackage{amsmath}

\begin{document}
\newcommand{\msun}{M$_\odot$}
\newcommand{\lsun}{L$_\odot$}
\newcommand{\mbh}{M$_\mathrm{BH}$}
\newcommand{\rhogas}{$\rho_{\mathrm{gas}}$}
\newcommand{\cs}{c$_s$}
\newcommand{\mbondi}{$\dot{\mathrm{M}}_{\mathrm{Bondi}}$}
\newcommand{\mdotbh}{$\dot{\mathrm{M}}_{\mathrm{BH}}$}
\newcommand\sbullet[1][.5]{\mathbin{\vcenter{\hbox{\scalebox{#1}{$\bullet$}}}}}

\title{Direct Tests of Black Hole Accretion Rate Prescriptions: I. Bondi Accretion at Different Scales}
\author[0000-0002-8122-3032]{James Agostino}
\affiliation{Ritter Astrophysical Research Center and Department of Physics \& Astronomy, University of Toledo, Toledo, OH 43606, USA}
\author[0000-0001-7264-0902]{Ming-Yi Lin}
\affiliation{Ritter Astrophysical Research Center and Department of Physics \& Astronomy, University of Toledo, Toledo, OH 43606, USA}
\author{Natasha Jones}
\affiliation{Ritter Astrophysical Research Center and Department of Physics \& Astronomy, University of Toledo, Toledo, OH 43606, USA}
\author[0000-0001-7421-2944]{Anne M. Medling}
\affiliation{Ritter Astrophysical Research Center and Department of Physics \& Astronomy, University of Toledo, Toledo, OH 43606, USA}
\author[0000-0003-0057-8892]{Loreto Barcos-Mu\~{n}oz} 
\affiliation{National Radio Astronomy Observatory, 520 Edgemont Road, Charlottesville, VA, 22903, USA}
\affiliation{Department of Astronomy, University of Virginia, 530 McCormick Road, Charlottesville, VA, 22903, USA}
\author[0000-0001-5769-4945]{Daniel Angl\'{e}s-Alc\'{a}zar}
\affiliation{Department of Physics, University of Connecticut, 196 Auditorium Road, U-3046, Storrs, CT 06269, USA}
\author[0000-0001-5231-2645]{Claudio Ricci}
\affiliation{Instituto de Estudios Astrof\'{\i}sicos, Facultad de Ingenier\'{\i}a y Ciencias, Universidad Diego Portales, Avenida Ejercito Libertador 441, Santiago, Chile}
\affiliation{Kavli Institute for Astronomy and Astrophysics, Peking University, Beijing 100871, China}
\author[0000-0003-3474-1125]{George C. Privon}
\affiliation{National Radio Astronomy Observatory, 520 Edgemont Road, Charlottesville, VA 22903}
\affiliation{Department of Astronomy, University of Florida, P.O. Box 112055, Gainesville, FL 32611, USA}
\affiliation{Department of Astronomy, University of Virginia, 530 McCormick Road, Charlottesville, VA 22904, USA}
\author[0000-0002-1912-0024]{Vivian U}
\affiliation{4129 Frederick Reines Hall, Department of Physics and Astronomy, University of California, Irvine, CA 92697, USA}
\affiliation{IPAC, California Institute of Technology, 1200 East California Boulevard, Pasadena, CA 91125, USA}
\author[0000-0002-5653-0786]{Paul Torrey}
\affiliation{Department of Astronomy, University of Virginia, 530 McCormick Road, Charlottesville, VA 22904}
\affiliation{Virginia Institute for Theoretical Astronomy, University of Virginia, Charlottesville, VA 22904, USA}
\affiliation{The NSF-Simons AI Institute for Cosmic Origins, USA}
\author[0000-0003-3729-1684]{Philip F. Hopkins}
\affiliation{TAPIR, Mailcode 350-17, California Institute of Technology, Pasadena, CA 91125, USA}
\author[0000-0003-0682-5436]{Claire Max}
\affiliation{Department of Astronomy \& Astrophysics, University of California, Santa Cruz, CA 95064, USA}

\onecolumngrid

\begin{abstract}
We present spatially resolved parsec-scale measurements of nuclear conditions (gas density and kinetic temperature) relevant for black hole accretion rate predictions in the Seyfert 2 galaxy, NGC 1068. We inject these parameters into the prescription for a Bondi-like accretion model, then compare the resulting accretion rate prediction to the empirical accretion rate derived from hard X-ray observations. Cosmological simulations have spatial resolution ranging from $\sim$10 pc to $\sim$kpc scales, and so for reasonable comparison we test these accretion rate predictions in pixel-sized radial steps out to 500 pc. Compared to warm H$_2$ gas, CO gas is the dominant mass carrier close to the SMBH. We find that the Bondi accretion rate (\mbondi) of cold molecular gas alone (measured using CO) overestimates the true accretion rate by up to 14 dex in a small aperture (r$\lesssim$5 pc) around the black hole, and by at least 8 dex inside large apertures (r$\lesssim$500 pc). These results are the first in a series of direct tests of accretion rate prescriptions, and they suggest that using a Bondi accretion formalism to model supermassive black hole accretion in Seyfert 2 galaxies may lead to overestimated accretion rates in simulations.
\end{abstract}

\section{Introduction}

Supermassive black holes (SMBHs), despite their small gravitational radius of the sphere of influence (1$\sim$100 pc), are thought to be a key piece of the connection between pc and kpc scales of galaxy evolution. Observations of galaxies with active galactic nuclei (AGN) have shown both directly and indirectly that AGN can inject energy into their surrounding environments, which can ultimately quench or in some cases trigger star formation (see \citealt{Fabian2012} for an observational review; \cite{Mercedes2023} for a simulated example). 
\label{sec:intro}
AGN are not just important to the central part of galaxies, they may also significantly impact several global properties of galaxies and their surrounding inter-galactic media. Relationships between black hole mass and global galaxy properties, like the velocity dispersion of stars in the galactic bulge, have been well-calibrated and show tight correlations (see \citealt{Kormendy2013}; \citealt{McConnell2013} for reviews). These correlations suggest that AGN radiative feedback, which in part depends on black hole mass, may leave an imprint on bulge stellar velocity dispersion (see \citealt{Ferrarese2000}; \citealt{Gebhardt2000} for seminal studies) but fueling regulation (\citealt{Escala2007}; \citealt{Chen2013}; \citealt{Angles2013}; \citealt{Angles2017}) and non-causal mass averaging through mergers (\citealt{Peng2007}; \citealt{Hirschmann2010}; \citealt{Knud2011}) have also been proposed as plausible drivers of black hole-galaxy scaling relations. Star formation in massive halos is suppressed (e.g. in \citealt{Behroozi2013}; \citealt{Torrey2014}), which could be caused by heating of the interstellar medium (ISM) from AGN feedback. In the high energy regime, a discrepancy is found between the observed and expected correlations between X-ray luminosities and temperatures of gas in the intra-cluster medium (called the L$_X$-T relation, see \citealt{Mushotzky1984}; \citealt{Markevitch1998}). This discrepancy suggests that gas in the intra-cluster medium evolves differently from dark matter;  energetics input by host AGN could be a factor as to why. 

Alongside indirect cases of the impact of AGN feedback on galaxy formation histories, the direct effects of AGN on the ISM have been observed for decades. Since more than 100 years ago (M87; \citealt{Curtis1918}) radio jets powered by a central SMBH have been seen to extend up to $\sim$0.9 Mpc outside from their host galaxies (e.g. Centaurus A; \citealt{Burns1983}). Outflows driven by these SMBHs have been observed in the process of depleting the ISM at outflow rates of 700 \msun\ yr$^{-1}$ (e.g. in Mrk 231; \citealt{Feruglio2010}). NGC 1068, which is the test case in the rest of this paper, has a complex and well studied AGN-driven outflow that has been observed to impact its ISM on sub-kpc scales (e.g.\citealt{Wilson1983}; \citealt{Muller2011}; \citealt{Garcia2014}; \citealt{Saito2022}; \citealt{Hviding2023}; \citealt{Holden2023}; \citealt{Gallimore2023}; \citealt{Mutie2024}; \citealt{Hagiwara2024}).

The energy output of an AGN is driven by mass accretion onto its accretion disk, fueled by inflows in the nuclei of galaxies. This gas then accelerates to speeds of up to $>0.1$c in the accretion disk, and that disk can power radiative outflows. These energetics interact with the ISM, the effects of which we call feedback. The direct observational feedback can be classified as two mechanisms: radiative (quasar mode) or kinematic (radio mode) (\citealt{Fabian2012}). In the quasar mode, occurring when the black hole accretes mass quickly, photons from the accretion disk couple to the ISM, transferring momentum in a powerful jet. In radio mode, accretion onto the disk is slower, and the primary feedback mechanism is in the form of collimated radio jets that typically appear narrower than quasar-mode jets (see \citealt{Cielo18} for a simulated comparison between the feedback of the two modes). Both modes can drive outflows, but the quasar-mode is thought to start the quenching process (the spatial extent of which grows over time) and then the radio-mode maintains that quenched state (see \citealt{Fabian2012} and \citealt{Morganti2017} for reviews). 


As simulations of galaxy evolution have been informed by increasingly detailed observations, theorists have begun to study the physical mechanisms that drive AGN feedback and how that feedback impacts the simulated host galaxies. \cite{Dubois2013} (see also \citealt{Dubois2013a}; \citealt{Taylor2015}) examined how AGN jets impact cold gas and transform blue, disky galaxies into red ellipticals. Building on these studies, \cite{Rosas2015}, who simulated accretion in galaxies of varied halo mass, find that in galaxies with M$_{\mathrm{Halo}}$ above 10$^{11.5}$\ \msun, as was observed in \cite{Behroozi2013}, star formation is suppressed by AGN feedback. \cite{Valentini2020} perform a suite of cosmological simulations in which they couple AGN feedback to different phases of the ISM. They find, in part, that energy output from the AGN as feedback must couple with both the cold and hot phases in order to avoid excessive SMBH growth.

As is seen in both observations and simulations, global galaxy properties can be affected by accretion-dependent feedback. Theorists have attempted to model the physical processes causing those properties to change. \cite{Williamson2020} perform radiation hydrodynamics modeling of the 1-100 pc scales in a nuclear region of a simulated AGN host. They demonstrate that increasingly polar winds are produced when anisotropic radiation from the AGN shifts the mass distribution of the outflow originating from the AGN. \cite{Meenakshi2022} simulated the direct interaction between AGN jet-induced outflows on 2 kpc scale and the ISM and found shocked emission fronts in the ISM that could be responsible for stunting star formation. On r $<$ 1 pc scale, \cite{Wada2023} were able to induce radiation-driven dusty outflows which impact the ISM as they continue on their outward paths. Tying the small and large scales together has been an ongoing challenge. 

Due to computational constraints, large-scale cosmological simulations that can model hundreds of Mpc$^3$ at a time are not able to directly resolve the physical processes that drive gas accretion at $<<$ 1 pc scales where accretion takes place, and so sub-grid prescriptions for black hole accretion and its subsequent feedback must be adopted.\footnote{Although there is also much variation in AGN feedback prescriptions, this program will focus on discussing the accretion rate prescriptions, on which all feedback depends.} The `sub-grid' is defined as the region below the gridded resolution of the simulation. Unfortunately, there is no unified model for these sub-grid physics, and different studies use different accretion prescriptions. The most commonly applied prescription is the one described in \cite{Bondi1952}, often referred to as the Bondi accretion prescription. The equation for the mass accretion rate follows the form: 

\begin{equation}
\label{eqn:mbondibase}
\dot{\mathrm{M}}_{\mathrm{Bondi}} = \frac{4\pi \mathrm{G}^2\mathrm{M^2_{BH}}\rho}{\mathrm{(c^2_{s} + v^2_{rel})}^{3/2}}
\end{equation}

where G is the gravitational constant, \mbh\ is the mass of the black hole, $\rho$ is the gas density, c$_s$ is the sound speed, and $v_{rel}$ is the relative velocity of the gas. In the pure Bondi case, the gas is assumed to be stationary relative to the galactic potential, so $v_{rel}$ is zero. This model is theoretically predicated on gas free-falling onto the SMBH once it reaches the Bondi radius, R$_{\mathrm{Bondi}}$ = 2G\mbh/$c_s^2$. The simulated Bondi radius is where the escape velocity of the SMBH (based on its mass) equals the sound speed of the gas in the nuclear region. The physical scale of the Bondi radius is typically of order 0.1-300 pc if we assume c$_\mathrm{s}$ of 400 km s$^{-1}$ (for hot gas) and SMBH mass range of 10$^6\sim10^9$ \msun. Some large scale cosmological simulation suites use a pure Bondi prescription to account for SMBH accretion, including \textit{MassiveBlack-II} (\citealt{Khandai2015}) and \textit{IllustrisTNG} (\citealt{Weinberger2017}; \citealt{Pillepich2018}).  

The Bondi accretion formalism, based on modern observations, is not a physically accurate picture of accretion as it ignores mechanisms such as angular momentum, outflows, and magnetic fields. Ignoring angular momentum of the gas and interactions due to self-gravity between the gas, stellar, and dark matter components, is only appropriate in the case of hot, virialized gas (\citealt{Hobbs2012}; \citealt{Negri2017}; \citealt{Angles2021}). Recent studies have shown that gas and other accreting material still has angular momentum inside what may be the Bondi radius, particularly in gas-rich mergers or galaxies with Seyfert AGN  (e.g. in \citealt{Davies2004}; \citealt{Hicks2013}; \citealt{Medling2014}; \citealt{Lin2016}), and so Bondi accretion timescales may be much shorter than in reality where angular momentum delays accretion. Feedback from the AGN in such models self-regulates this rapid growth (\citealt{Angles2015}). 

Feedback in the form of outflows may also be a factor in the accretion picture. As we discuss further in Section~\ref{sec:structuredescription}, NGC 1068 hosts a complex of outflows that may be serving as a regulator for black hole growth (e.g. \citealt{Wilson1983}; \citealt{Muller2011}; \citealt{Garcia2014}; \citealt{Saito2022}; \citealt{Hviding2023}; \citealt{Holden2023}; \citealt{Gallimore2023}; \citealt{Mutie2024}; \citealt{Hagiwara2024}). Gas available for accretion in NGC 1068-like galaxies may be heated and/or driven away from the SMBH on scales as small as 2 pc (\citealt{Gallimore2016}), well within the Bondi radius inside which accretion is assumed to happen. \cite{Garcia2014}, for example, estimate the AGN-driven outflow rate in NGC 1068's circum-nuclear disk (CND; r $\sim$ 200 pc) to be 63 \msun\ yr$^{-1}$. Whether the origin of outflows in the nuclei of galaxies is driven primarily by an AGN or star formation, their feedback can regulate SMBH accretion.

Magnetic fields are also an integral part of most detailed models of AGN fueling and feedback (e.g. \citealt{Blandford1977}; \citealt{Blandford1977}; \cite{Tchekhovskoy2011}; \citealt{Ghisellini2014}; \citealt{Zamaninasab2014}; \citealt{Davis2020}; \cite{Narayan2022}), but are not part of Bondi accretion. \cite{Lopez2020} found magnetically aligned dust grains in on the 4 pc (torus) scale in NGC 1068. \cite{Gallimore2023} find evidence for ordered magnetic fields on even smaller scales NGC 1068's complex of water masers. They mention that these maser filaments qualitatively match the outflow features of the \cite{Emmering1992} magnetic accretion disk model. Magnetic fields have been shown play a role in facilitating accretion and feedback, and do seem to be present in NGC 1068's nucleus. Angular momentum, outflows, and magnetic fields are all thought to play an important part in fueling a SMBH, but are not part of the Bondi accretion formalism.

Because of the M$^2 _\mathrm{BH}$ dependence of accretion rate in Bondi accretion prescriptions, low mass BH seed growth is suppressed such that BHs do not grow quick enough to match their expected mass at corresponding redshifts. To account for this discrepancy, some large-scale cosmological simulation suites adjust the accretion physics by using modified versions of Bondi accretion. The prescription in the \textit{Illustris} (the predecessor to \textit{IllustrisTNG}; \citealt{Vogelsberger2013}; \citealt{Illustris2014}) and \textit{Magneticum Pathfinder} hydrodynamical simulation suites (\citealt{Hirschmann2014}; \citealt{Bocquet2016}; \citealt{Dolag2016}) modify Bondi by multiplying Equation~\ref{eqn:mbondibase} by a constant (unitless) `boost' factor $\alpha$ (following the prescription of \citealt{Springel2005}; \citealt{DiMatteo2005}; \citealt{SpringelHernquist2005}). The boost factor is used to account for the volume average of the Bondi-rates for both the cold and hot phases in the simulations and typically has a value = 100. Another large-scale cosmological model, \textit{Horizon-AGN} (\citealt{Dubois2016}), uses an $\alpha$ similar to \textit{Illustris} and \textit{Magneticum}, but instead of a constant value, their boost factor (following the prescription from \citealt{Booth2009}; see also \citealt{Dubois2012}) is $\alpha = (\rho / \rho_0)^2$ or $\alpha = 1$ for densities above and below the threshold for star formation respectively. \textit{EAGLE} (\citealt{Eagle2015}) uses a pure Bondi prescription alongside the heuristic correction from \cite{Rosas2015} to account for variable angular momentum of accreting gas. Another approach, used by the large-scale \textit{Romulus} suite (\citealt{Tremmel2017}) is to adjust the Bondi accretion rate depending on the motion of the simulated gas particles. In \textit{Romulus}, if the smallest relative velocity (which they equate to v$_{\mathrm{bulk}}$, the bulk motion of the gas) of the gas particle closest to the SMBH is faster than the rotational velocity of the gas, they replace the relative velocity of the SMBH (in Equation~\ref{eqn:mbondibase}) with v$_{\mathrm{bulk}}$ and multiply the Bondi rate by a density-dependent boost factor similar to \textit{Horizon-AGN}.



Bondi or Bondi-like accretion prescriptions are the most commonly used, but theorists have also designed accretion prescriptions with very different underlying physics. One large-scale simulation ([100 $h^{-1}$ Mpc]$^3$ volume) suite that in part uses one of these prescriptions is \textit{SIMBA} (\citealt{Dave2019}). In \textit{SIMBA}, pure Bondi accretion is still applied for hot gas accretion where, as we mentioned, it is most appropriate. But, they then apply a torque-limited accretion formalism for the cold gas where instabilities in the disk drive mass inflow (\citealt{Hopkins2011}; \citealt{Angles2017}). 

Understanding if and in which cases different sub-grid prescriptions are accurately estimating accretion rates onto the black holes of galaxies is critically important to cosmological simulations and conclusions drawn from them. Without an accurate prescription for accretion over time, simulations cannot accurately implement the impact of AGN feedback, and as such may have incorrect outcomes with regards to galaxy formation and evolution. Depending on the assumed accretion prescription, simulations find that BH scaling relations are driven either by feedback efficiency (in Bondi-like models), or accretion efficiency (in a torque-driven model; see \citealt{Angles2021} for further discussion). Theorists' conclusions on which physics drive the co-evolution between BH mass and global galaxy properties is directly dependent on which accretion model is implemented. Determining which accretion formalism is most appropriate in which circumstances is critical to understanding BH-galaxy co-evolution in our Universe.


Observationally testing how black hole accretion rate prescriptions perform has only become possible in recent times. In this study, in the prototypical Seyfert 2 galaxy NGC 1068, we directly measure the parameters that go into Bondi accretion, \rhogas\ and \cs,  on physical scales ranging from 0.1-500 pc -- inside the sub-grid regime for the aforementioned simulations. To achieve the high resolution required for the measurements we use observations of the cold and warm components of the nuclear gas from ALMA mm interferometry and Keck/OSIRIS NIR integral field spectroscopy (see Section~\ref{sec:data}). We then plug these measured parameters into the pure Bondi accretion prescription as a function of radius to mimic what a simulation at that resolution would estimate for the black hole accretion rate. Finally, we test these predicted Bondi accretion rates against empirically derived accretion rates using hard (14-195 keV) X-ray data from The Burst Alert Telescope (BAT) AGN Spectroscopic Survey (BASS) (\citealt{Ricci2017}). The BAT instrument (\citealt{Barthelmy2005}; \citealt{Krimm2013}) on \textit{Swift} (\citealt{Gehrels2004}) is a hard X-ray detector that surveys the entire sky, reporting X-ray sources to within 1-4 arcmin accuracy.


\begin{figure*}
    \centering    \includegraphics[width=0.550\linewidth]{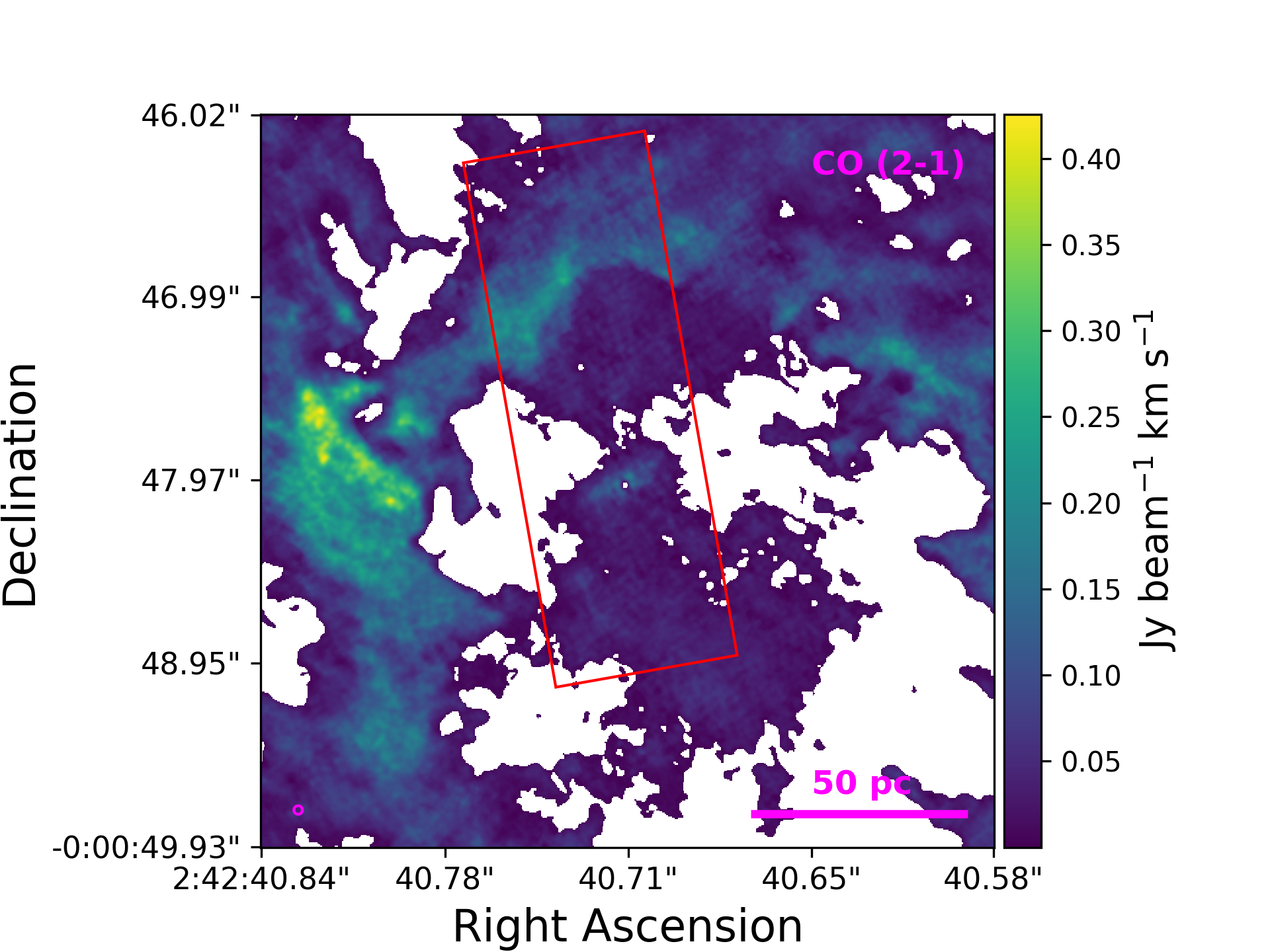}   \includegraphics[width=0.1850\linewidth]{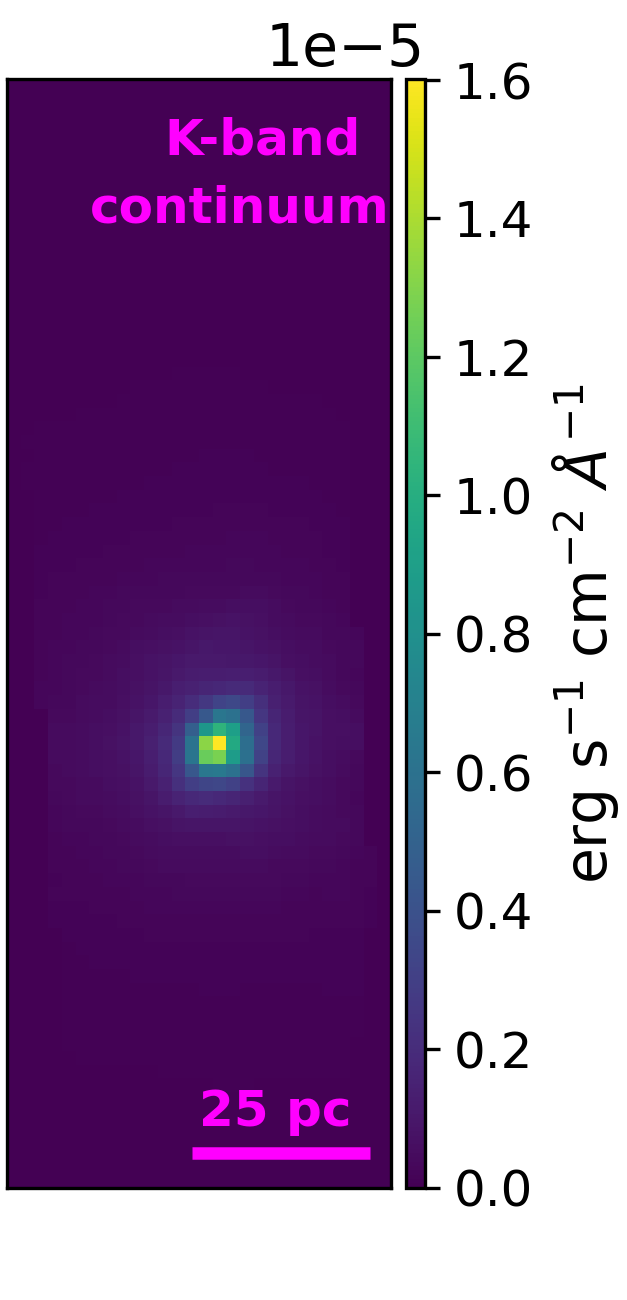}   \includegraphics[width=0.1700\linewidth]{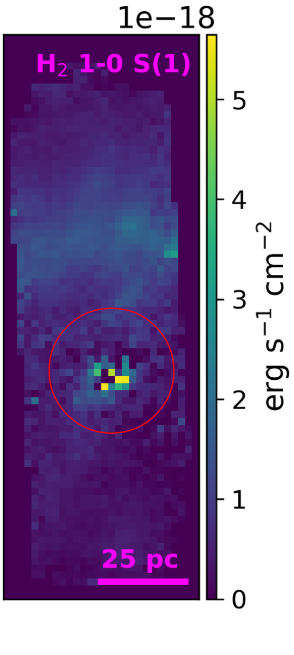}
    \caption{Nuclear region of NGC 1068 in the CO(2-1) flux (left, from ALMA), 2.2$\mu$m line+continuum (middle), and the continuum subtracted rovibrational H$_2$ 1-0 S(1) transition (right), described in Section~\ref{sec:data}. The CO(2-1) moment 0 map is masked below 3$\times$rms and the red box in the CO(2-1) moment map represents the field of view of the NIR data. The very small, magenta circle in the bottom left represents the beam size of the ALMA data (20$\times$20 mas). All three images show flux peaks at the AGN's location and both the CO and H$_2$ maps have enhanced emission in the CND ring. The red circle in the H$_2$ 1-0 S(1) moment map represents the aperture in which T$_\mathrm{kin}$ is calculated in Figure~\ref{fig:excitation}.}
    \label{fig:morphology}
\end{figure*}

In this work, we use cosmological parameters of H$_0$ = 70 km~s$^{-1}$ Mpc$^{-1}$, $\Omega$$_m$ = 0.28, and $\Omega$$_\Lambda$ = 0.72 (\citealt{2009ApJS..180..225H}). To calculate spatial scales and luminosity distance to NGC 1068 we use Ned Wright's Cosmology Calculator (\citealt{Wright2006}).

\section{Nuclear structure of NGC 1068}
\label{sec:structuredescription}

NGC 1068 is one of the most-studied prototypical Seyfert galaxies, and as such a wealth of information has already been published about its nuclear structure. The studies described here are not an exhaustive list, but are included to provide context relevant to our analysis.

At 2.2 pc resolution, NGC 1068 hosts a complex of water masers that are thought to originate from the accretion disk on much smaller ($<$0.1 pc) scales. \cite{Greenhill1996} observed the masers with very long baseline interferometry (VLBI) using both the Very Long Baseline Array and Very Large Array to achieve 0.65 pc resolution. They used the velocity gradient of the maser emission to infer a rotational velocity of the gas, and in turn constrain \mbh $\sim 1 \times10^7$ \msun. \cite{Kumar1999} modeled the 0.65-1.1 pc disk from which the maser emission is thought to be ejected from. The clumps in their disk model interact with each other, leading to eventual accretion onto the SMBH. More recently, \cite{Gallimore2023} used High Sensitivity Array observations of the water masers to determine its position to 0.3 mas precision. Using these data, they find that the disk kinematics are `consistent with Keplerian motion' where a central mass (presumably the SMBH) dominates the dynamics of the maser disk. Using these findings they infer a central mass of (1.7 $\pm$ 0.01) $\times$ 10$^7$ \msun within $\sim$7 pc.

On slightly larger scales with near and mid infrared interferometry, multiple authors were able to resolve a two-component dusty torus (\citealt{Jaffe2004}; \citealt{Raban2009}). One component is smaller and more elongated (1.35 $\times$ 0.45 pc) in size than the other (3 $\times$ 4 pc). In the nucleus of Circinus, another Seyfert 2 AGN, \cite{Tristram2014} also found a two-component dusty torus. Images like these that showed structure inconsistent with the prior, observationally-defined, Type 2 classification of these galaxies (unless foreground extinction was applied) fundamentally challenged the AGN unification model (\citealt{Antonucci1993}). 

\cite{Gamez2022} used sub-pc resolution observations of NGC 1068 taken with the MATISSE/ESO/VLTI interferometer between 3 and 13 $\mu$m to map the dust temperature distribution of the dust observed in the previously mentioned studies. They confirm an optically thick pc scale dusty structure and a second, less optically thick disk that extends to at least 10 pc. \cite{Garcia2019} (who in part use the same ALMA data as we describe in Section~\ref{sec:submmdata}) find a 14 pc CO(2-1) nuclear disk with a PA ($\sim$110-140 deg) aligned with the water maser disk PA (140 deg). Also in \cite{Garcia2019}, they observe the circumnuclear disk (CND), which as can be seen in Figure~\ref{fig:morphology}, has a gas deficit inside the CND in its central $\sim$130 pc region.

To resolve the kinematics of the 10 pc inner disk (often referred to as the torus) and CND, \cite{Impellizzeri2019} observed HCN J=(3-2) with ALMA at 1.4 pc resolution (project code 2017.1.01666.S; PI Gallimore). They found a pc scale outflow signature in the HCN absorption profile corresponding to a velocity of $\sim$450 km s$^{-1}$. These are faster even than the $\sim$400 km s$^{-1}$ velocities in the bipolar outflow that \cite{Gallimore2016} found in CO(6-5). \cite{Impellizzeri2019} and \cite{Imanishi2020} find that the torus, as observed with these dense gas tracers, rotates in the opposite direction with respect to the CND and water maser emission. This is particularly surprising because the water maser emission is rotating in the same direction as the CND rather than the molecular torus it is physically closer to. In \cite{Garcia2019}, the authors also find counter-rotation in CO(2-1). They find that a `significant part' of the observed counter-rotation in CO(2-1) can be attributed to a northern AGN-driven wind. To make a more robust determination on the northern wind's impact on the counter-rotation, \cite{Garcia2019} say that higher resolution data is required so that the outflowing component can be better disentangled from the rotating component.

Outflows originating from the AGN as highlighted in the previous paragraph can serve to regulate black hole accretion. NGC 1068 hosts a complex outflow in the NE direction, perpendicular to the nuclear disk. The largest outflow component is a kpc-scale radio jet (e.g. in \citealt{Gallimore1996a}). \cite{Mutie2024} present higher resolution ($\sim$4 pc) $e$-MERLIN 5 GHz data along with archival VLA 10 GHz, and VLA 21 GHz images of the jet. These images together show not only the central jet emission, but also detail in the larger scale bow shock, $>$200 pc from the SMBH in the same NE direction, which exhibits direct evidence of the AGN's impact on the ISM. The impact of the jet on the ISM is studied in part in both \cite{Hviding2023} and \cite{Holden2023}, who both show evidence for gas ionization consistent with shock ionization or radiation-bounded AGN-photoionization along the outflow's path on 160 pc to kpc scale. \cite{Huang2022} examined the properties of the shock tracers SiO and HNCO in five regions of NGC 1068's nucleus, finding that SiO traces the fast shocks and HNCO traces the slow or non-shocked regions (see also \citealt{Kelly2017}). \cite{Garcia2014} show that the CO kinematics on distances 50 to 400 pc are spatially correlated with the radio jet, evidence that the AGN is influencing even the cold ISM. ALMA CO(6-5) observations from \cite{Gallimore2016} show that this molecular outflow originates within 2 pc from the SMBH, and has velocities relative to systemic of about 400 km s$^{-1}$. \cite{Impellizzeri2019} used the \cite{Gallimore2016} Band 9 CO data together with HCN (J = 3-2) to show that the CO absorption reveals a P-Cigny profile, indicative of an outflow. NGC 1068 hosts multiple outflow components which may impact the SMBH's ability to consume additional mass.

\section{NGC 1068 observations}
\label{sec:data}

For NGC 1068, we made use of $<$3 pc scale resolution observations both in the near infrared (NIR) with Keck/OSIRIS+AO (adaptive optics; PI Medling), and in the mm with ALMA archival data (PI Garc{\'i}a-Burillo). High resolution observations like these are critical to radially sampling the predicted Bondi accretion rate in Section~\ref{sec:prescriptions}.  

\subsection{Keck/OSIRIS K-band Integral Field Spectroscopy}
\label{sec:NIRdata}
The first of two sets of data we are using in this project is a set of high resolution integral field unit (IFU) Keck/OSIRIS+AO (OH-Suppressing InfraRed Imaging Spectrograph, \citealt{OSIRIS}) integrations, for which we mosaic all frames into a single data cube. These observations were taken with the Kbb filter (broad-band K between 1.965 - 2.381 $\mu$m) with the 35 mas pixel$^{-1}$ plate scale on 2018 December 28th, 2019 January 22nd, and 2019 October 7th for a total exposure time of 6120 seconds (51 frames, 120 seconds each). Weather impacted observations on 2019 October 7th, during which the laser guiding system was also not working. For NGC 1068 we used the galaxy nucleus as the natural guide star in NGS mode, and as the tip/tilt star in LGS mode. AO corrections in those frames without the laser produced larger point spread functions with full-width at half-maximum (FWHM) values between 3 and 5 pixels compared to $\sim$2 with the laser on other nights. We reduced the Keck/OSIRIS+AO observations using the OSIRIS Data Reduction Pipeline (OSIRISDRP, \citealt{Lyke2017}; \citealt{Lockhart2019}) version 4.2.0, which we use to extract a spectrum for each spatial pixel, assemble the spectra into a cube, and mosaic the 51 total frames together to form the final image, which has a 0.17" point spread function (PSF) FWHM. Flux calibration was applied for each night before final mosaicking.

The resulting mosaic reveals a strong K-Band continuum (particularly near the AGN) and H$_2$ 1-0 rovibrational emission (S(0), $\lambda_{\mathrm{rest}} = 2.2235\mu \mathrm{m}$;  S(1), $\lambda_{\mathrm{rest}} = 2.1218\mu \mathrm{m}$; S(2)), $\lambda_{\mathrm{rest}} = 2.0338\mu \mathrm{m}$. These line+continuum and continuum-subtracted H$_2$ 1-0 S(1) maps are shown in the middle and right panels of Figure~\ref{fig:morphology} respectively. The line+continuum map was made using the Cube Analysis and Rendering Tool for Astronomy (\verb|CARTA|, \citealt{CARTA2021}) and the continuum subtracted H$_2$ 1-0 S(1) map was made using \verb|QFitsView| (\citealt{qfitsview}). Both images show peaks of emission on or near the position of the central engine, and NGC 1068's CND ring can be seen in the H$_2$ map.

\subsection{ALMA Band 6 Long baseline Interferometry}
\label{sec:submmdata}

We chose the highest resolution CO J =(2-1) (hereafter CO(2-1)) available on the ALMA archive (PI Gallimore, project code 2017.1.01666.S). Prof. Gallimore shared these data with us via private communication. The continuum subtracted spectral cube has a rms of 0.13 mJy over 10 km s$^{-1}$, and was imaged as a multi-scale clean using scales of 0, 5, and 15 pixels. NGC 1068 was imaged using a Briggs (\citealt{Briggs1995}) robust value of 0, resulting in a beam size of 20 $\times$ 20 mas. For our final calculations we used the moment 0 (flux) map of the CO(2-1) emission also provided by Prof. Gallimore. Figure~\ref{fig:morphology} (left) shows this CO(2-1) moment 0 map which is masked below 3$\times$rms and is used for our analysis in Section~\ref{sec:prescriptions}. Like in the warm H$_2$ observations, the CND ring is a bright source in CO(2-1).

\section{Prescription parameters}
\label{sec:prescriptions}
In this pilot study, we examine the performance of the most commonly used accretion prescription for black hole growth in NGC 1068. The Bondi accretion formalism with a relative velocity of zero (also known as Bondi-Hoyle, or Bondi-Hoyle-Lyttleton e.g. \citealt{Hoyle1939}; \citealt{Hoyle1944}; \citealt{Bondi1952})  follows the form: 
\begin{equation}
\label{eqn:mbondi}
\dot{\mathrm{M}}_{\mathrm{Bondi}} = 4\pi \mathrm{G}^2\mathrm{M^2_{BH}}\rho\mathrm{c_s}^{-3}
\end{equation}
where G is the gravitational constant, \mbh\ is the mass of the black hole, $\rho$ is the gas density and c$_s$ is the sound speed. 

Bondi accretion is predicated on a spherically symmetric, non-self-gravitating gas distribution in which the gas inside the Bondi radius has no angular momentum, does not have an outflowing components, and is not shaped by magnetic field lines. While this kind of environment may not be most appropriate for describing all galaxy nuclei (see Section~\ref{sec:meaning} for additional information), Bondi accretion is a simple analytical prescription that can be applied inside a sphere of any radius, which makes it convenient as a sub-grid prescription. 

In this subsection, we outline the methods for measuring each free parameter in the Bondi prescription using the available high resolution data from Section~\ref{sec:data}.
\subsection{Parameter 1: black hole mass}
\label{sec:mbh}

\cite{Greenhill1996} imaged NGC 1068's water maser emission at 0.65 pc scales using very long baseline interferometry. From the rotation curve of the water maser emission, they found the enclosed mass within that radius to be $\sim$1 $\times$ 10$^7$ \msun (with uncertainty of order unity). Another study by \cite{Lodato2003} derive a smaller black hole mass of $\sim$(8 $\pm$0.3) $\times$ 10$^6$ in a self-gravitating accretion disk model that matches the \cite{Greenhill1996} and \cite{Greenhill1997} observations well. The \cite{Lodato2003} model corrects for non-Keplerian motion in the velocity profile of the water maser emission, but this could be an over-correction. In fact, other studies have found that the disk rotation may still be dominated by the black hole (\citealt{Imanishi2018}). Albeit with a worse fit to the velocities from the maser emission, \cite{Lodato2003} also fit a Keplerian rotation model, which has a best fit black hole mass of $\sim$(1.5 $\pm$0.02) $\times$ 10$^7$ \msun. \cite{Gallimore2023}, using maser emission observed with the High Sensitivity Array (which pinpoints the SMBH's location to a 0.3 mas precision) and a Keplerian rotation model, find a very similar value of (1.7 $\pm$ 0.01) $\times$ 10$^7$ \msun. We input the \cite{Gallimore2023} value throughout our analysis as it is the most precise measurement of NGC 1068's SMBH mass. 

\subsection{Parameter 2: gas density}
\label{sec:gasdensity}

\begin{figure*}
    \centering
    \includegraphics[width=0.45\linewidth]{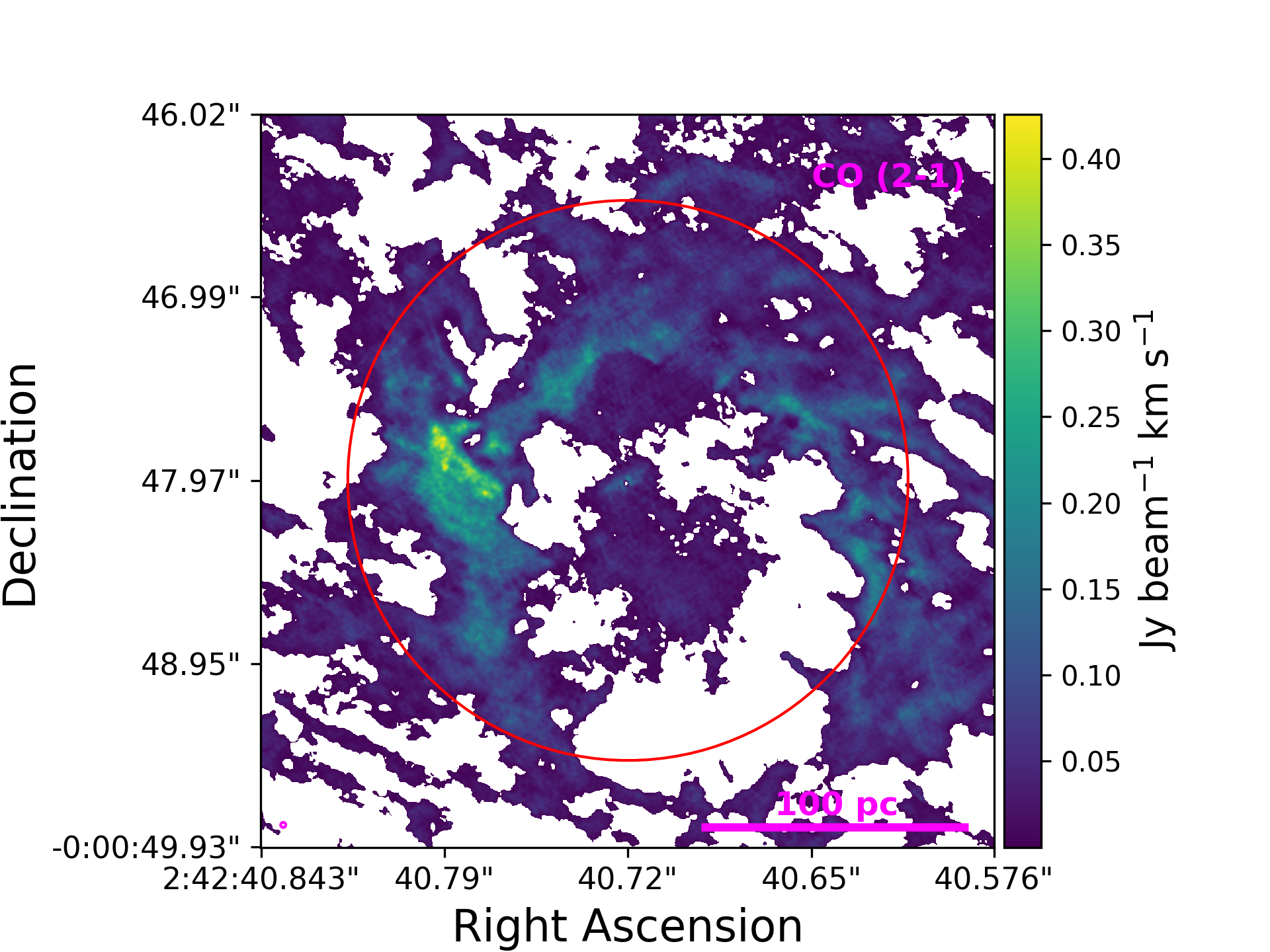}
    \includegraphics[width=0.45\linewidth]{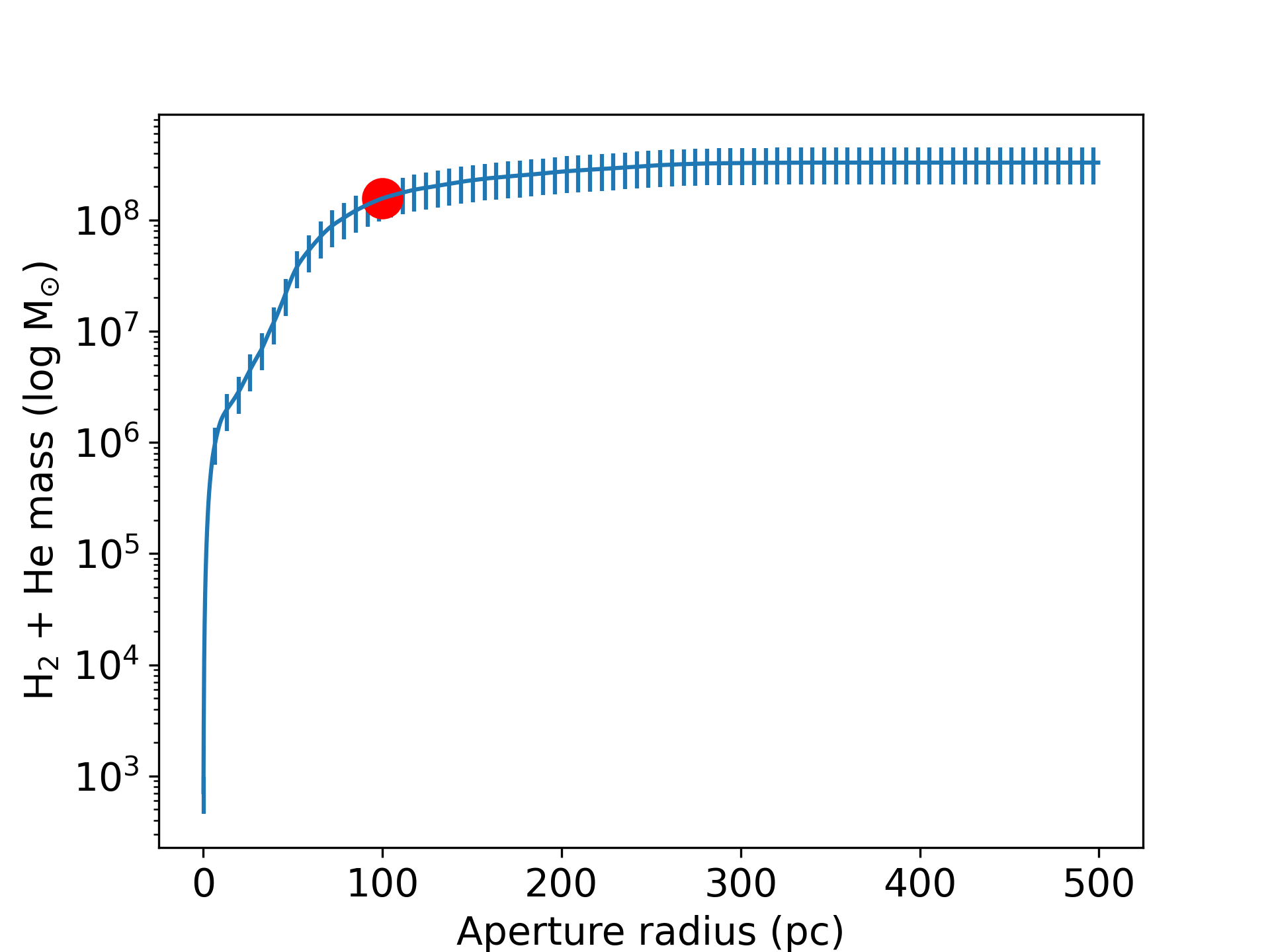}
     \caption{\textit{Left}: Moment 0 map of CO(2-1) in NGC 1068, with flux density values masked below 3$\times$rms. The red circle indicates the location and size of the 100 pc aperture (centered on the AGN) corresponding to the red dot in the right panel, which has M$_{\mathrm{enc, H_2+He}}$ = (1.57 $\pm\ 0.57) \times$10$^8$\ \msun. The small, magenta circle in the bottom left represents the beam size of the ALMA data 20$\times$20 mas). \textit{Right}: Integrated mass profile inside the radial aperture. Details on the conversion to molecular gas mass can be found in Section~\ref{sec:gasdensity}.}
    \label{fig:menclosed}
\end{figure*}
\subsubsection{Choice of volume element}
\label{sec:volumeelement}
To measure the gas density, we first must define our volume element. In cosmological simulations, typically, a fixed number of gas particles exist inside a spherical region with radius r centered on the location of the SMBH. This volume makes up the black hole kernel, in which the accretion physics are prescribed. Although studies like the ones discussed in Section~\ref{sec:structuredescription} and \cite{Vollmer2022} have shown that the $\sim$10 pc cold gas distribution is more disk-like, we opt to use a sphere of volume V =$\frac{4}{3}\pi \mathrm{r}^3$ centered on the AGN for which we vary the radius with the goal of mimicking the spherical radial aperture that simulations typically use to evaluate Bondi accretion.

\subsubsection{Cold gas mass}
\label{sec:coldgasmass}
To measure the cold molecular gas (H$_2$ and He) mass inside the sphere, we use the CO(2-1) data described in Section~\ref{sec:submmdata}. To obtain a molecular gas mass, we utilize the conversion factor $\alpha _{\mathrm{CO}}$. The exact value of $\alpha _{\mathrm{CO}}$ depends on several factors including the size scale and environment over which the CO flux is measured. The picture is further complicated by the distinction between $\alpha _{\mathrm{CO(1-0)}}$ and $\alpha _{\mathrm{CO(2-1)}}$, where the difference is dictated by the ratio between the line luminosity of the two rotational transitions: r$_{21}$ (r$_{21}$ = $\mathrm{L}'_{\mathrm{CO(2-1)}}$/$\mathrm{L}'_\mathrm{CO(1-0)}$), which depends on the temperature of the gas. In this work, we follow the same $\alpha _{\mathrm{CO}}$ methodology as in \cite{Garcia2019} who use the Milky Way $\alpha _{\mathrm{CO(1-0)}}$ = 4.3 $\pm$ 1.29 M$_{\odot}
(\mathrm{K \ km \ s}^{-1} \mathrm{pc}^2)^{-1}$ recommended by \cite{Bolatto2013}. We use $\alpha _{\mathrm{CO(1-0)}}$ in conjunction with the averaged line intensity ratios for NGC 1068's northern and southern CND regions (because the CND ring contains the majority of the nuclear gas mass): r$_{21}$ = 2.2 $\pm$ 0.4,  from \cite{Viti2014} to calculate a final 
\begin{multline}
\label{eqn:alphaco}
\alpha_{\mathrm{CO(2-1)}} = \frac{\alpha_{\mathrm{CO(1-0)}}}{\mathrm{r}_{21}} 
=  \frac{4.3 \ \pm \ 1.29}{2.2 \ \pm 0.4}  \\ = 1.95 \ \pm \ 0.73 \ M_{\odot}
(\mathrm{K \ km \ s}^{-1} \ \mathrm{pc}^2)^{-1}.
\end{multline}
The outflowing components of NGC 1068 may have a lower $\alpha _{\mathrm{CO(1-0)}}$, but we expect the Milky Way value to be closer to the average for the purpose of measuring integrated enclosed masses, especially at larger r. The outflow's impact on our gas mass measurement is expected to be small as there is not much CO(2-1) emission between the AGN and CND ring, and the CND ring itself does not visually appear disturbed along the path of the outflow. $\alpha_{\mathrm{CO(2-1)}}$ is then multiplied by the luminosity inside a circular aperture of radius r, to match our spherical geometry. The enclosed mass profile is shown alongside a snapshot of the aperture geometry in Figure~\ref{fig:menclosed}. 

\cite{Garcia2019}, who center their r = 200 pc aperture measurement on the center of the CND ring, find a molecular (H$_2$ + helium) gas mass of $\simeq$1.4 $\times$ 10$^8$\ \msun. We measure molecular gas mass within the same aperture by measuring the CO(2-1) line flux using a matched \verb|CARTA| region on the CO(2-1) line intensity map (See Figure~\ref{fig:morphology}, left) and converting to \msun\ units using Equation~\ref{eqn:alphaco}. We find (2.70 $\pm$ 0.99) $\times$ 10$^8$\ \msun \, slightly higher than the \cite{Garcia2019} estimate.

To convert enclosed mass to density we divide by the volume of the sphere with radius r (see Section~\ref{sec:volumeelement}) with r defined by our circular aperture size used for measuring mass. In this sphere with r = 100 pc centered on the AGN as shown in Figure~\ref{fig:menclosed} (left), we find a molecular gas mass density of 37.39 $\pm$ 13.65 \msun pc$^{-3}$. 
\subsubsection{Warm H$_2$ gas mass}
\label{sec:warm gas mass}
We also calculate an enclosed mass using the warm ($\sim 2000 $K; see \citealt{Scoville1982}, \citealt{Riffel2008}) H$_2$ gas measured from the NIR data, following Equation 6 of \cite{Storchi2009}, which uses the line flux of the H$_2$ 1-0 S(1) rovibrational transition at $\lambda_{\mathrm{rest}}$ = 2.1218 $\mu$m. \cite{Martins2010} used the NASA 3-m Infrared Telescope Facility (IRTF) to observe NGC 1068 and found a nuclear (slit 1"x 2") extinction E(B-V) of 1.13 (from their Table 4). Assuming the standard extinction law of \cite{Cardelli1989} with R$_\mathrm{v}$ = 3.1, the extinction A$_\mathrm{v}$ (A$_\mathrm{v}$ = R$_\mathrm{v}$ $\times$ E(B-V)) is $\sim$3.5. Based on A$_\mathrm{k}$$\sim$A$_\mathrm{v}$/10 (\citealt{Howarth1983}), we measure the H$_2$ 1-0 S(1) extinction-corrected intrinsic flux (F$_{\mathrm{intrinsic}}$ = F$_{\mathrm{observed}} \times 10^{(0.4A_\mathrm{k})})$ and directly convert it to the warm H$_2$ gas mass. The extinction-corrected H$_2$ gas mass inside r $<$1.7" (111 pc) is $\sim$68 \msun, about 1.38 times the (uncorrected) observed value. Due to the rectangular FoV, only an aperture radius of $<$0.3" is fully contained within the OSIRIS FOV, suggesting that H$_2$ emission at radii $>$0.3" is incomplete. Regardless, the warm H$_2$ mass is inconsequential compared to the CO-derived value of (2.50 $\pm\ 0.92) \times$10$^7$\ \msun\ in the same rectangular region as the OSIRIS FoV. 
 
Other than the field of view, a primary reason that the warm gas measurement in this region may be so small is due to the radiative environment in NGC 1068's nucleus. Under local thermal equilibrium (LTE, where the energy distribution can be described by a single number locally) conditions, the H$_2$ emission can be excited by the equilibrium value for temperatures T $\approx$ 1000 K (\citealt{Davies2005}). To reach such high excitation temperatures in NGC 1068, H$_2$ emission lines can be excited through several mechanisms, as described below:

(1) \textit{UV fluorescence}: This excitation mechanism dominates in photodissociation regions (PDRs). Far-ultraviolet (FUV, $\lambda$ $>$ 912 \AA) radiation pumps the molecules into electronically excited states, leading to subsequent cascades that emit fluorescent emission (\citealt{Wakelam2017}). \cite{Davies2005} presented near-infrared spectra of the circumnuclear regions of nearby Seyfert 1 galaxies and found that the v = 1 level is fully thermalized, while higher vibrational levels (v = 2 and v = 3) exhibit a suppressed ortho-para ratio, indicating fluorescent excitation. According to the unified model of AGN, the intrinsic excitation mechanisms in Seyfert 1 and Seyfert 2 galaxies are expected to be the same, with the only difference arising from the observer’s viewing angle. Although NGC 1068 is classified as a Seyfert 2 galaxy and is therefore expected to detect weaker FUV radiation from the observers, the HST/Faint Object Camera (FOC) UV imaging still reveals a polarized nucleus (\citealt{Barnouin2023}) within our Keck/OSIRIS field of view (see Section~\ref{sec:NIRdata}). 

(2) \textit{Shocks and outflows}: \cite{Veilleux1997} suggest that shocks associated with nuclear outflows are a likely heating source for H$_2$ in many Seyfert 2 galaxies. These effects have been observed directly in NGC 1068.
\cite{May2017} analyzed VLT/SINFONI and Gemini/NIFS data with a larger FOV covering the entire CND and proposed that the CND could be an expanding bubble. As discussed in Section~\ref{sec:structuredescription}, works such as \cite{Gallimore2016} and \cite{Impellizzeri2019} have observed $\gtrsim$400 km s$^{-1}$ molecular outflows that start on the pc scale and expand to the kpc scale (e.g. in \citealt{Mutie2024}). These outflows have been shown to shock the CND (e.g. in \citealt{Huang2022}; \citealt{Hviding2023}; and \citealt{Holden2023}), which can drive high local temperatures.

(3) \textit{X-ray heating from the AGN}:  X-ray emission can penetrate deeply into regions that are opaque to UV photons and influence H$_2$ excitation (\citealt{Matt1997}). In NGC 1068 and other galaxies, the ISM has been shown to host altered chemical abundances of molecular gas typically attributed to X-ray radiation (e.g. \citealt{Usero2004}; \citealt{Garcia2010}; \citealt{Izumi2016}; \citealt{Butterworth2024}). All of these mechanisms can contribute to measured H$_2$ emission.


\subsection{Parameter 3: sound speed of the gas}
\label{sec:soundspeed}
The final parameter required in the Bondi accretion formalism is the sound speed of the gas. The sound speed for an ideal gas is:
\begin{equation}
    c_s = \sqrt{\frac{\gamma k_B T_K}{\mu m_p}}
    \label{eqn:cs}
\end{equation}

where $\gamma$ is the adiabatic index (1, as the gas is assumed to be isothermal in each sub-region), k$_B$ is the Boltzmann constant 1.381 $\times$ 10$^{-16}$ erg $^{-1}$ K$^{-1}$, T$_K$ is the temperature of the gas (K), and $\mu$ is the mean molecular weight of the gas, which is 2.7 since we assume the molecular gas is H$_2$, 10\% helium, and trace metals, and m$_p$ is the mass of a proton (kg). All but the temperature in this case are constants. 

For the temperature of the molecular gas, we use two methods: one using CO rotation diagrams (cold gas), and another using an excitation diagram for the molecular H$_2$ (warm gas) from our Keck/OSIRIS+AO NIR data. 

\subsubsection{CO-derived c$_{\mathrm{s}}$}
For a temperature from CO transitions we refer to the work of \cite{Viti2014} who infer the temperature of the gas in the CND of NGC 1068 by using CO rotation diagrams. This method assumes that  the gas is in LTE, and that the observations are mostly in the Rayleigh-Jeans regime where the intensity of the radiation is proportional to the temperature. This temperature is also known as the `rotational temperature' and is equal to the kinetic temperature if all CO levels are thermalized (\citealt{Goldsmith1999}). Because of these assumptions, this temperature should be considered a lower limit, which translates to an upper limit on our final accretion rate because \mbondi\ $\propto$ c$_\mathrm{s}^{-3}$. For the central region of NGC 1068, \cite{Viti2014} find a temperature of 50 $\pm$ 5-7 K via the CO rotation diagram method (see Section 3.1.1. of their work for more details). Plugging 50 $\pm$ 5-7 K and the other constants into Equation~\ref{eqn:cs}, we find that the speed of sound in the cold molecular gas phase is 391.0 $\pm$ 23.5 m s$^{-1}$. We note that the data used in the \cite{Viti2014} diagram ranges from 21 to 109 pc in its angular resolution. The data we use in this work has an angular resolution of $<$3 pc. We expect that the sound speed calculated in this paragraph is more reliable on the larger scales, e.g. in the CND, due to this resolution mismatch. Updated CO rotation diagrams using the higher resolution data now available on the ALMA archive (like the data used in this work) would be required to make more robust conclusions on the speed of sound in the cold molecular gas phase closer to the AGN.

\subsubsection{H$_2$-derived c$_{\mathrm{s}}$}
\label{sec:H2temp}
\begin{figure*}
    \centering
    \includegraphics[width=0.52\linewidth]{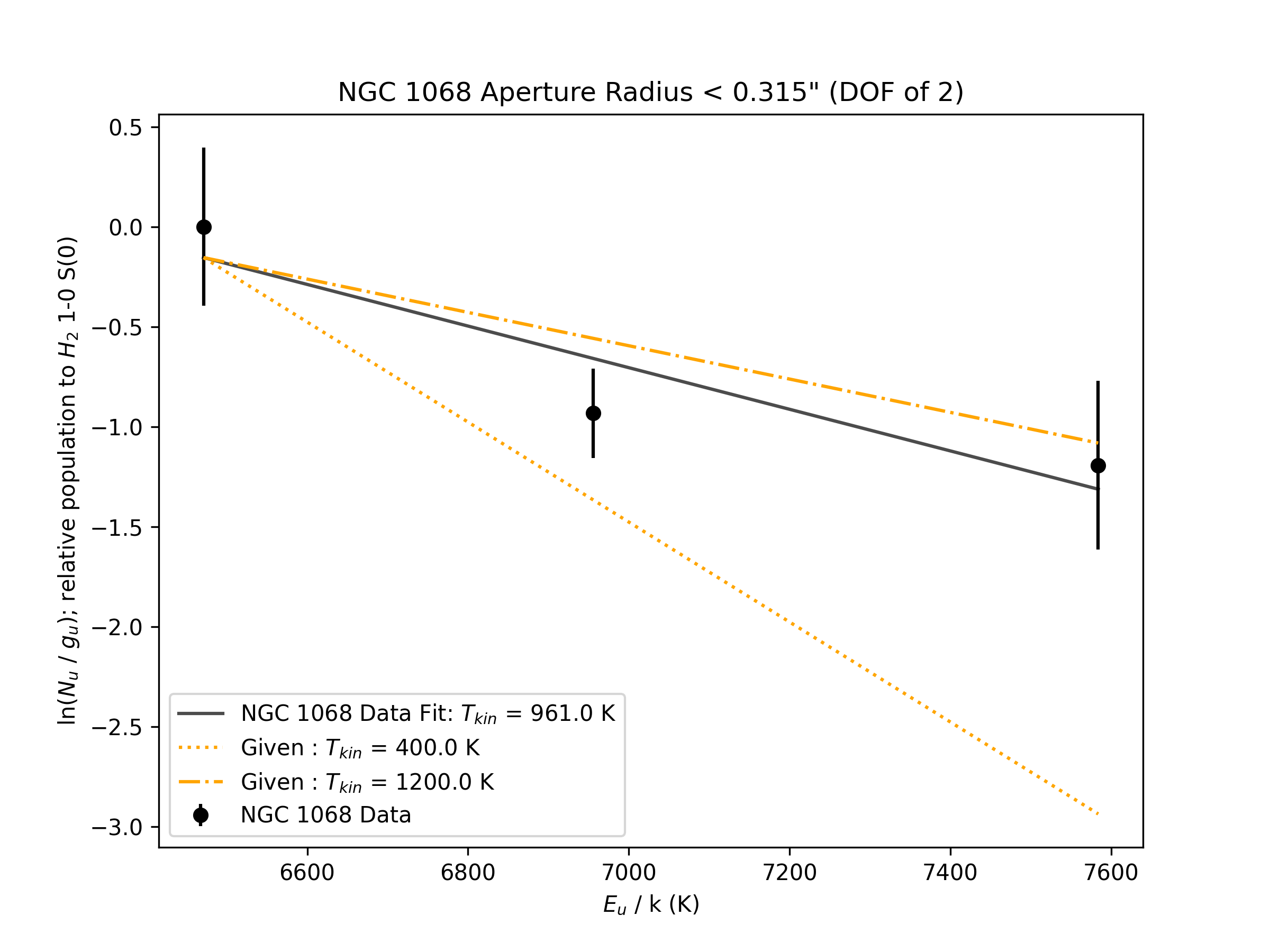}
    \includegraphics[width=0.45\linewidth]{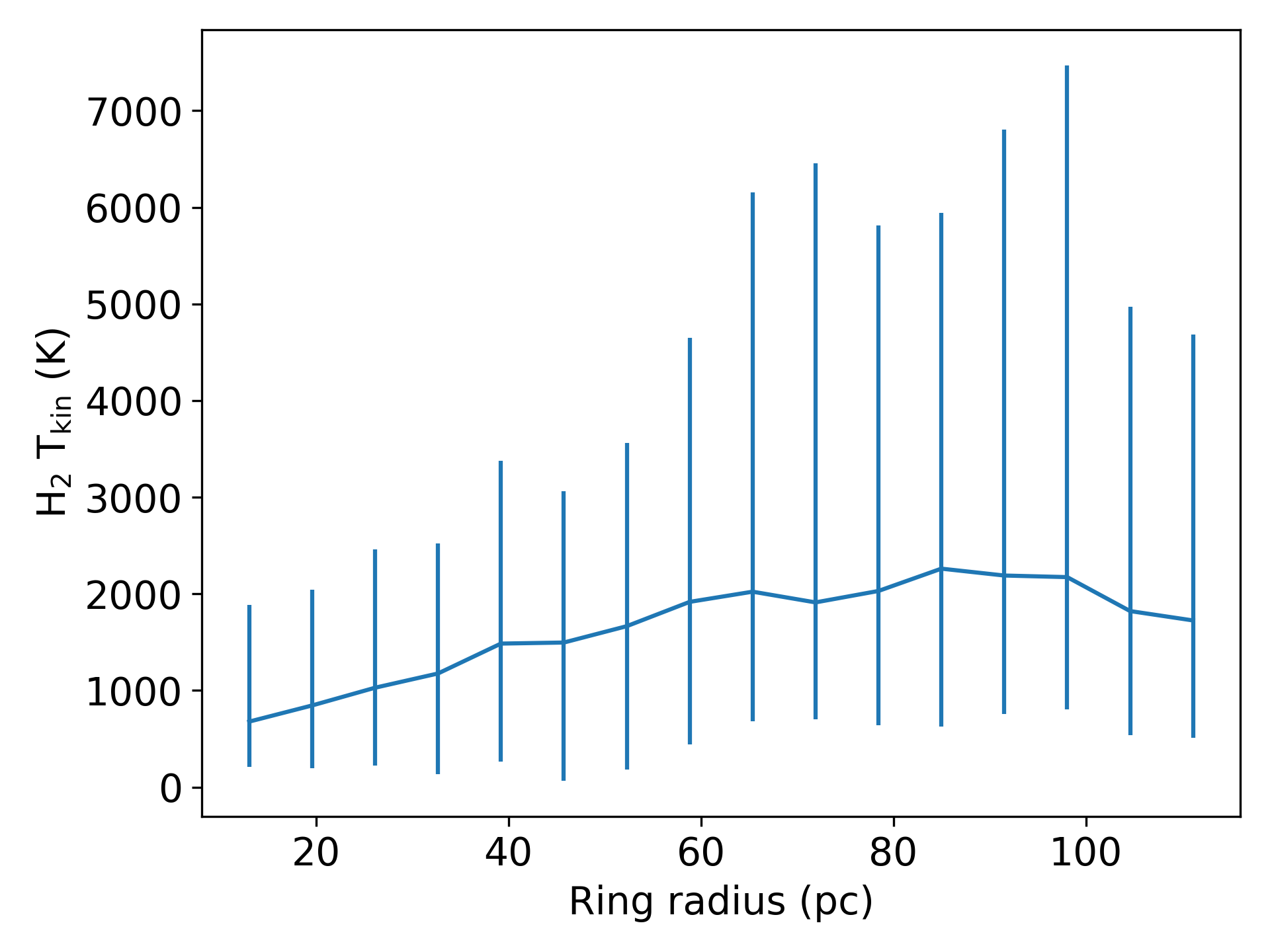}

    \caption{(\textit{Left}:) Column density in the upper level of each H$_2$ 1-0 S(0), S(1), and S(2) transition as a function of energy level of that transition as a temperature (K) inside a r = 21 pc circular aperture centered on the AGN as shown in Figure~\ref{fig:morphology}. The best fit slope (using linear regression), as described in Section~\ref{sec:H2temp}, is the temperature of the gas in that region if we assume LTE. (\textit{Right}:) T$_\mathrm{kin}$ estimated as in the excitation diagram on the left but instead inside circular apertures matching the methods of Section~\ref{sec:gasdensity} from 0.2" to 1.7" in steps of 0.1". The mean of the derived temperatures is 1612$\substack{+2840\\-1216}$ K.}
    \label{fig:excitation}
\end{figure*}

As shown in Section~\ref{sec:warm gas mass}, warm H$_2$ is also present in NGC 1068's nuclear regions, so we also consider the sound speed for this component of the ISM. To measure the temperature which we then use in the \cs\ calculation, we use the H$_2$ 1-0 S(0), S(1), and S(2) rovibrational line fluxes in the Keck/OSIRIS NIR data described in Section~\ref{sec:NIRdata}. Assuming the H$_2$ gas is in LTE, the H$_2$ excitation temperature is equal to the kinetic temperature. Figure~\ref{fig:excitation}(a) shows the H$_2$ excitation diagram, which is the column density in the upper level of each transition normalized by its statistical weight ($N_u / g_u$) as a function of energy of the level as a temperature ($E_u$). The best-fit slope of this relationship is related to $T_K$ as $\frac{N_u}{g_u} \propto e^{({-\frac{h\nu}{kT_K}})}$ in the LTE description of energy level populations (see pages 322, 327 of \citealt{Wilson2013}). Solving for $T_K$ then yields $-\frac{1}{T_K} \propto \frac{ln\frac{N_u}{g_u}}{\frac{E_u}{k}}$. 

Because we have spatially resolved data for these H$_2$ lines, we can derive kinetic temperatures from 12-111 pc and apply them at the matched distances in the accretion rate prediction. While the Keck/OSIRIS+AO data has a higher resolution than 6 pc, the H$_2$ 1-0 S(1) and S(2) lines are not detected in a r$\le$6 pc (0.1") aperture. Fluxes of the lines are measured using the line fitting tool in \verb|QFitsView| (\citealt{qfitsview}), which we use to fit the continuum and one Gaussian component to the integrated (within a region circular region with radius r) spectrum. Figure~\ref{fig:excitation}(b) shows the range of excitation temperatures as a function of radius. T$_\mathrm{kin}$ ranges from 678-2261 K, and peaks at r$\le$85 pc where T$_\mathrm{kin}$ = 2261 $ \substack{+3683\\-1631}$ K. High temperatures may be caused by the influence of the PDR (Section~\ref{sec:coldgasmass} describes observations of this for NGC 1068), which is found to increase the H$_2$ 1-0 S(1) emission by up to 70$\%$ in the some luminous infrared galaxies (\citealt{Davies2000Mrk266}; \citealt{Davies2003ULIRGs}). Using Equation~\ref{eqn:cs} (with a mean molecular weight of H$_2$ only) results in warm H$_2$ sound speeds between 1440-2629 m s$^{-1}$, peaking at r = 85 pc.  

\section{Results: \mbondi vs. $\dot{\mathrm{M}}_{\mathrm{X-ray}}$}
\label{sec:output}

\subsection{Calculating \mbondi}
\begin{figure*}
    \centering
    \includegraphics[width=1\linewidth]{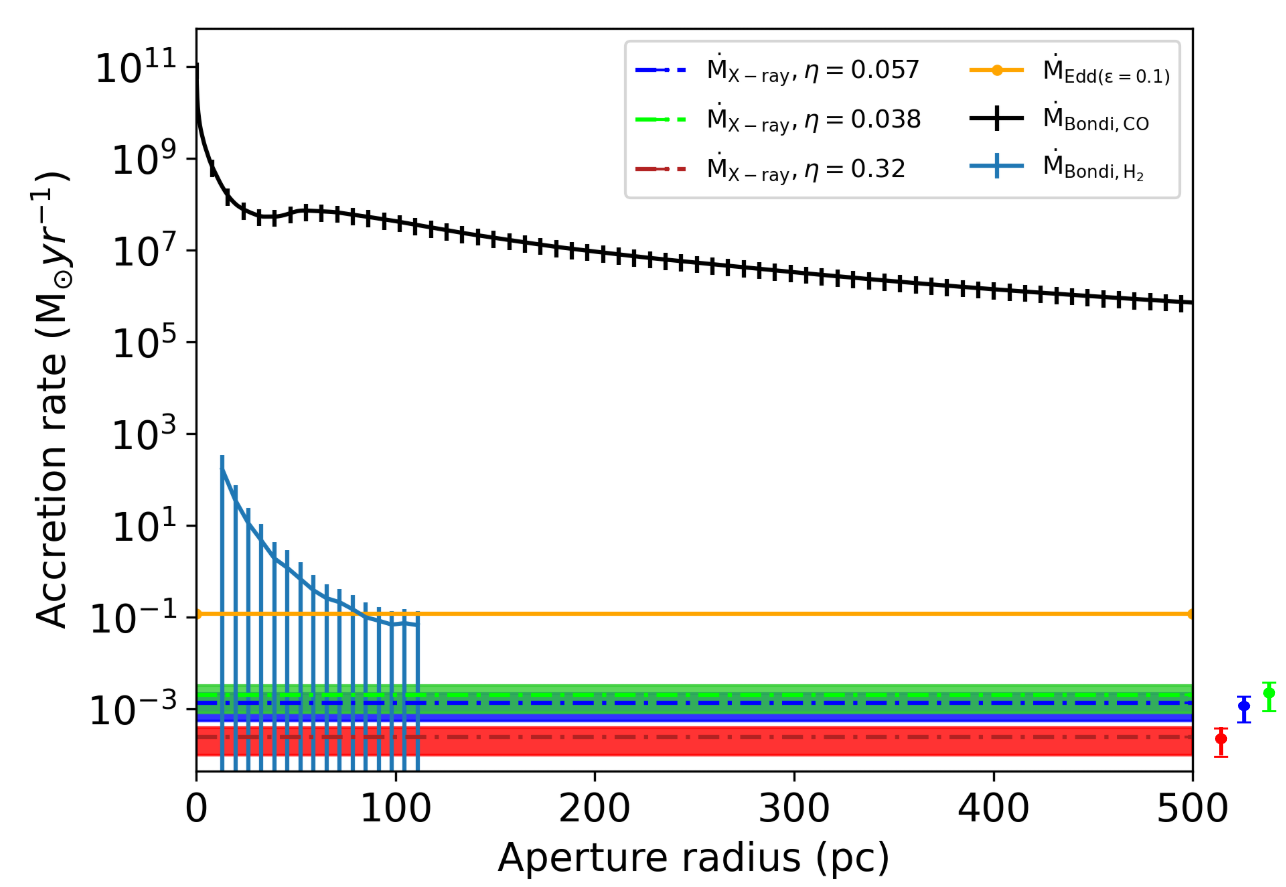}
    \caption{X-ray and (Cold-derived) Bondi accretion rates as a function of radius over which nuclear parameters are measured. Color bars and ranges displayed to the right denote uncertainties on each X-ray measurement. Regardless of which gas component is used to estimate \mbondi, the Bondi prescription overestimates \mdotbh\ by orders of magnitude, and is above the Eddington rate (orange line, with radiative efficiency $\epsilon$ = 0.1). For the cold gas case, which represents the majority of gas available for accretion in NGC 1068, Bondi overpredicts the accretion rate by between 10 and 14 orders of magnitude.}
    \label{fig:bondixray}
\end{figure*}

Now that we have calculated each parameter for the Bondi accretion prescription in Section~\ref{sec:prescriptions}, we are ready to estimate a Bondi accretion rate. Because our parameters are spatially resolved, we calculate accretion rate as a function of radial distance r representing a simulated resolution: 
\begin{equation}
\label{eqn:mbondir}
\dot{\mathrm{M}}_{\mathrm{Bondi}} (r) = 4\pi \mathrm{G}^2\mathrm{M^2_{BH}}\rho(\le r)\mathrm{c_s(\le r)^{-3}} .
\end{equation}
where \mbh\ is in kg, $\rho$ is in kg m$^{-3}$ and c$_\mathrm{s}$ is in m s$^{-1}$. 
Figure~\ref{fig:bondixray} shows the Bondi accretion rate for the cold and warm derived cases as a function of radius. The Bondi accretion rate derived from the cold gas component ranges between about 10$^{6}$ \msun\ yr$^{-1}$ and 10$^{10}$ \msun\ yr$^{-1}$. As the enclosed mass found in Section~\ref{sec:gasdensity} for the warm H$_2$ gas component in r$<$ 170 pc is small (68 \msun), and the temperature gradient is high (678-2261 K, see Section~\ref{sec:H2temp}) relative to the values found for the cold gas component, the resulting Bondi accretion rates are much smaller (between about 10$^{-2}$ and 11 \msun\ yr$^{-1}$) for the warm gas. These results suggest that the cold gas is the dominant carrier of mass accretion on r$<$ 170 pc scales within the Bondi framework. Table~\ref{table:accretionrates} shows a range of precise values for both the cold and warm Bondi accretion rates.

\subsection{Calculating X-ray accretion rates}
\label{sec:comparison}
To understand how well the Bondi accretion formalism compares to the real accretion rate, we compare it to the X-ray derived accretion rate. To calculate an accretion rate from X-ray measurements, we use \textit{Swift}/BAT data from the BAT AGN Spectroscopic Survey (BASS, \citealt{Ricci2017}). They present intrinsic luminosities in the 14-195 keV band, which we use alongside the bolometric correction, Equation 17 in \cite{Gupta2024}:
\begin{multline}    
\label{eqn:bolocorrection}
\mathrm{log}(\kappa_{14-195}) = (0.13 \pm 0.04) \times \mathrm{log}(\lambda_{\mathrm{Edd}}) \\
+ (1.04 \pm 0.05) 
\end{multline}

\noindent to calculate bolometric luminosity. Because \cite{Ricci2017} measure a neutral column density of logN$_\mathrm{H}$ = 25.0 cm$^{-2}$ in NGC 1068 and the X-ray continuum might not be well estimated when the emission is dominated by reprocessed radiation in environments like this, we conservatively estimate uncertainty on the input intrinsic 14-195 keV luminosity to be $\pm$ 0.4 dex (for insight into the mm and MIR to X-ray correlations that helped inform this choice, see \citealt{Asmus2015}, \citealt{Kawamuro2022}, and \citealt{Ricci2023}). We then use that bolometric luminosity in the equation from \cite{Netzer2014}, L$_{\mathrm{bol}}$ = $\eta\dot{\mathrm{M}}\mathrm{c}^2$, solving for $\dot{\mathrm{M}}$ where $\eta$ is the unitless mass-to-radiation conversion efficiency that depends on the spin of the black hole. For stationary, retrograde disk, and maximally rotating SMBHs respectively, the values for $\eta$ are 0.057, 0.038, and 0.32 (\citealt{Netzer2014}). For NGC 1068, we find $\dot{\mathrm{M}}{_{\mathrm{X-ray}}}$ values equal to (1.36 $\pm$ 0.81) $\times 10^{-3}$ \msun yr$^{-1}$ (stationary SMBH), (2.04  $\pm$ 1.21) $\times 10^{-3}$ \msun yr$^{-1}$ (retrograde accretion disk), and (2.42 $\pm$ 1.44) $\times 10^{-4}$ \msun yr$^{-1}$ (maximally spinning SMBH). As shown in Figure~\ref{fig:bondixray} and Table~\ref{table:accretionrates}, \mbondi\ overestimates the accretion rate by several orders of magnitude in the warm gas case to up to 14 orders of magnitude in the cold gas case in small aperture radii. In Section~\ref{sec:meaning} we discuss the implications of such a discrepancy with respect to cosmological simulations.


\cite{Vollmer2022} used the IR-derived bolometric luminosity for the AGN in NGC 1068 from \cite{Vollmer2018} to calculate \mdotbh $\sim$ L$_{bol}$/(0.1c$^2$) $\sim$ 0.05 \msun yr$^{-1}$. They calculate a mass accretion rate onto their modeled accretion disk for NGC 1068 to be 2 $\times$ 10$^{-3}$ \msun yr$^{-1}$ ($\eta = 0.1$), which is in agreement with our $\dot{\mathrm{M}}{_{\mathrm{X-ray}}}$ values.


\section{Discussion: results in the context of simulations}
\label{sec:meaning}
\begin{table*}[ht]
\centering
\scalebox{0.8}{
\begin{tabular}{l|c c c c c c c c}
         Method & \ & \ & \ & Accretion rate & (\msun\ yr$^{-1}$) & \ & \ \\
         \hline
          X-rays ($\epsilon$= 0.038) & \ & \ & \ & (2.04 $\pm$ 1.21)& $\times$ 10$^{-3}$ & \ \\
         X-rays ($\epsilon$= 0.057) & \ & \ & \ & (1.36 $\pm$ 0.81)& $\times$ 10$^{-3}$ & \ \\  
         X-rays ($\epsilon$= 0.32) & \ & \ & \ & (2.42 $\pm$ 1.44)& $\times$ 10$^{-4}$ & \ \\
         \hline
         \hline
         Bondi ($<$r) & 5 pc & 10 pc & 25 pc & 50 pc & 100 pc & 200 pc &  500 pc \\
         \hline
         (T$_{\mathrm{Kin}}$ = 50 K) & 1.53 $\pm$ & 4.31 $\pm$ & 7.31 $\pm$ & 6.98 $\pm$ & 4.30 $\pm$ & 9.36 $\pm$ & 7.24 $\pm$\\
         \ & 0.63 $\times10^{9}$ & 1.76 $\times10^{8}$ & 2.98 $\times10^{7}$ & 2.84 $\times10^{7}$  & 1.75 $\times10^{7}$ & 3.81 $\times10^{6}$ & 
 2.93 $\times10^{5}$ \\

          (T$_{\mathrm{Kin}}$ = 678-2261 K) & $^a$ & $^a$ & 1.12 $\substack{+2.33 \\ -1.31}$ & 0.68 $\substack{+1.16 \\ -0.91}$ & 6.91 $\substack{+25.2 \\ -6.52}$ & $^a$ & $^a$ \\
         \ & \ & \ & $\times 10^{1}$ & \ & $\times 10^{-2}$ & \ & \ \\         
    \end{tabular}}
    \caption{Accretion rate measurements and estimates for rates derived from X-ray luminosities and the pure Bondi accretion prescription inside various radii. In NGC 1068, where the cold gas phase makes up the bulk of the gas mass in the nucleus, the cold gas derived Bondi accretion estimate outpaces the X-ray derived accretion rates by between 10 and 14 orders of magnitude. $^a$ H$_2$ 1-0 S(1) and S(2) lines required to calculate temperature for the Bondi calculation are not detected in this aperture.}
    \label{table:accretionrates}
\end{table*}
To inform theorists on which accretion prescriptions in their simulations are best to use and when, we have designed our measurements to fit in the practical context of those simulations. Large scale cosmological simulations must use sub-grid physics for accretion because of computing constraints. As mentioned earlier, some examples of hydrodynamical galaxy evolution simulations that use spherically symmetric, Bondi or Bondi-like black hole accretion formalisms are \textit{Illustris/IllustrisTNG} (\citealt{Illustris2014}; \citealt{Vogelsberger2014};  \citealt{IllustrisTNG2018}), \textit{Magneticum Pathfinder} (\citealt{Hirschmann2014}; \citealt{Bocquet2016}; \citealt{Dolag2016}),  \textit{MassiveBlack-II} (\citealt{Khandai2015}), \textit{Eagle} (\citealt{Eagle2015}), \textit{Horizon-AGN} (\citealt{Dubois2016}), \textit{Romulus} (\citealt{Tremmel2017}), and \textit{SIMBA} (\citealt{SIMBA2019}, uses Bondi for hot gas only). The resolution of the hydrodynamical gas cells in which these sub-grid physics is typically close to 1 kpc. The properties of these gas cells is determined by two factors: mass, and temperature. In the highest resolution zoom-in simulations, the spherical radius in which particle calculations are made is approximately 10 pc (\citealt{Wetzel2023}). Hyper-refinement simulations (e.g. \citealt{Angles2021}; \citealt{Hopkins2024}), where gas resolution elements are dynamically split to reach high resolution can reach spatial scales smaller than 10 pc, but these simulations can only be practically run for short periods of cosmic time due to computing constraints.  

Because we have spatially-resolved measurements, we are able to examine the performance of Bondi accretion at a range of spatial scales. For the Bondi accretion rate derived from warm gas we are limited by the field of view of OSIRIS (0.56$\times$2.24" with our observational setup), but the ALMA data extends to over 500 pc away from the SMBH.

Table~\ref{table:accretionrates} shows the Bondi accretion rates estimated at radii between 5-500 pc as calculated in Section~\ref{sec:output}, and the X-ray accretion rates as calculated in Section~\ref{sec:comparison}, which are all plotted together in Figure~\ref{fig:bondixray}. At all aperture radii, regardless of whether we are estimating \mbondi\ using the cold or warm gas component, the parameterized Bondi accretion rate exceeds the X-ray derived accretion rate (by 1 or more dex in the warm gas case and by 9 or more dex in the cold gas case). 

This is, perhaps, not a surprising result. Past studies have hinted towards Bondi accretion overestimating the real accretion rate. \cite{DiMatteo2000} found that luminosities calculated using estimated Bondi accretion rates for six black holes with masses of 0.22-5.2 $\times$ 10$^9$ \msun\ determined in \cite{Magorrian1998} 
were 4-6 orders of magnitude higher than the real luminosities of the galaxy nuclei. \cite{Hopkins2016} model SMBH accretion in a gas-rich nuclear disk in a massive simulated galaxy with $0.1$ pc resolution. In their study, applying a pure Bondi accretion formalism resulted in an accretion rate $\sim$10$^8$ times higher than the luminosity-derived accretion rate native to their simulation. Their pure Bondi accretion rate ($\sim$$10^7$\msun yr$^{-1}$), agrees with our cold-gas derived pure Bondi accretion rate between approximately 25 and 200 pc in NGC 1068. Near the SMBH, pure Bondi accretion ignores the possibility that gas particles may have angular momentum. The gas in the simulation used in \cite{Hopkins2016} is primarily cold and is supported by angular momentum rather than thermal pressure. Observations show that especially in gas-rich galaxies that naturally host molecular torii, the r$<$100 pc cold gas reservoir is large, has significant angular momentum, and is the primary candidate for black hole accretion fueling (\citealt{Davies2004}; \citealt{Hicks2013}; \citealt{Medling2014}; \citealt{Lin2016}; \citealt{Gaspari2015}). Additionally, as discussed in Section~\ref{sec:intro}, outflows and magnetic fields (which are both apparent in NGC 1068) can be contributing factors to the SMBH fueling process.

Perhaps the most compelling reason for the overestimate of the Bondi accretion rate in this work compared to implementation in simulations can be traced back to the temperature of the gas in simulations. Large scale cosmological simulations do not have fine enough resolution to resolve the clumpy, cold, ISM. As a result, their gas particles are much hotter than the cold gas we find to be the primary mass carrier in NGC 1068. Simulations that use Bondi-like accretion such as \textit{Horizon-AGN} (\citealt{Dubois2012}; \citealt{Dubois2016}), \textit{IllustrisTNG} (\citealt{Weinberger2017}; \citealt{Pillepich2018}) and \textit{Massiveblack-II} (\citealt{Khandai2015}) find temperatures $\gtrsim 10^4$ K for their mass reservoirs in nuclear environments. High temperatures are also in part driven by feedback mechanisms. For example in \textit{Horizon-AGN} (\citealt{Dubois2012}; \citealt{Dubois2016}), quasar mode feedback from the AGN imparts energy only once the surrounding gas can be heated to 10$^7$ K. Radiative AGN feedback can also suppress atomic cooling (\citealt{Vogelsberger2013}). Such temperatures are much higher than the observed cold gas in NGC 1068's accretion reservoir and are likely too high for the majority of the real accreting media, but in simulations they suppress the Bondi accretion rate by a factor of T$^{-1.5}$, important for the regulation of BH growth.

If NGC 1068 is typical, in that the bulk of its gas available for accretion is cold, these results suggest that the usage of pure Bondi accretion is likely to struggle to accurately predict real black hole accretion rates. Our cold gas-derived Bondi accretion estimates predict dramatically high (by up to 14 orders of magnitude) accretion rates relative to the true, X-ray derived values.  Understanding the physical mechanisms that drive accretion on the sub-grid scales in galactic nuclei can inform the future development of accretion prescriptions.  The Bondi prescription allows particles to fall directly onto the BH inside the Bondi radius, but our results suggest that angular momentum, outflows, and magnetic fields play important roles in some nuclei.

\section{Conclusions and future expansion of this project}
\label{sec:conclusions}
In this study we estimate a Bondi accretion rate as a function of radius for NGC 1068 using two different molecular gas tracers resolved in the sub-grid regime, and compare the result to the direct accretion rate derived from hard X-ray luminosity of the AGN. Compared to warm H$_2$ gas, CO gas is the dominant mass carrier close to the SMBH. Following this, the cold gas derived Bondi accretion rate estimate outpaces the X-ray derived value by more than 8 orders of magnitude at all aperture sizes.

This paper is a pilot for a wider study of AGN and accretion prescriptions. Direct probes of sub-grid accretion prescriptions may, as our sample expands, help identify which physical processes dominate accretion on a variety of spatial scales, and in turn provide recommendations for appropriate sub-grid prescriptions to describe them. The results in this work support previous evidence that in high resolution cosmological simulations, applying a Bondi accretion prescription can lead to large overestimates of \mdotbh\ and therefore large overestimates of AGN feedback, which in turn impacts the global galaxy evolutionary track. We note that this is a test for a specific Seyfert 2 AGN. To make more robust recommendations about the application of the Bondi accretion prescription for sub-grid accretion physics we must directly test Bondi on more galaxies. 

\section{Acknowledgements}

The authors wish to recognize and acknowledge the very significant cultural role and reverence that the summit of Maunakea has always had within the indigenous Hawaiian community; we are privileged to be guests on your sacred mountain. We wish to pay respect to the Atacameño community of the Chajnantor
Plateau, whose traditional home now also includes the ALMA observatory. We thank Prof. Gallimore for sharing his excellent data products from this measurement set with us and thank Dr. Impellizzeri for connecting us with Prof. Gallimore. We also thank the anonymous referee, whose comments specifically regarding data usage and references have greatly improved this paper. This work makes use of the following data from ALMA: project 2017.1.01666.S (PI Gallimore).  ALMA is a partnership of ESO (representing its member states), NSF (USA) and NINS (Japan), together with NRC (Canada) and NSC and ASIAA (Taiwan) and KASI (Republic of Korea), in cooperation with the Republic of Chile. The Joint ALMA Observatory is operated by ESO, AUI/NRAO and NAOJ. The National Radio Astronomy Observatory is a facility of the National Science Foundation operated under cooperative agreement by Associated Universities, Inc. Some of the data presented herein were obtained at the W. M. Keck Observatory, which is operated as a scientific partnership among the California Institute of Technology, the University of California and the National Aeronautics and Space Administration. The Observatory was made possible by the generous financial support of the W. M. Keck Foundation. The authors also wish to thank the W.M. Keck Observatory staff for their efforts on the OSIRIS+AO instrumentation. JA, AMM, M-YL, and NJ acknowledge support from NSF CAREER grant number 2239807 and Cottrell Scholar Award CS-CSA-2024-092 from the Research Corporation for Science Advancement. PT acknowledges support from NSF-AST 2346977 and the NSF-Simons AI Institute for Cosmic Origins which is supported by the National Science Foundation under Cooperative Agreement 2421782 and the Simons Foundation award MPS-AI-00010515. D.A.A. acknowledges support from NSF grant AST-2108944 and CAREER award AST-2442788, NASA grant ATP23-0156, STScI grants JWST-GO-01712.009-A, JWST-AR-04357.001-A, and JWST-AR-05366.005-A, an Alfred P. Sloan Research Fellowship, and Cottrell Scholar Award CS-CSA-2023-028 by the Research Corporation for Science Advancement.

\textit{Software:} Astropy \citep{astropy:2013, astropy:2018, astropy:2022}, Matplotlib (\citealt{Matplotlib}), NumPy (\citealt{harris2020array}). 

\pagebreak
\bibliographystyle{apj}
\bibliography{prescriptions1citations}
\end{document}